\documentclass[structabstract]{aa}  
\usepackage{graphicx}
\usepackage{natbib}

\begin{document}
   \title{Updated Pre-Main Sequence tracks at low metallicities \\ for 0.1 $\le$ M/M$_{\odot}$$\le$ 1.5\thanks{These evolutionary tracks and isochrones are available in electronic form at a WEB site \texttt{http://www.mporzio.astro.it/$\%$7Etsa/}}}


   \author{M. Di Criscienzo \inst{1},
          P. Ventura \inst{1}
          \and           
          F. D'Antona \inst{1}
          }

   \institute{Osservatorio Astronomico di Roma,
              Via Frascati 33, 00040, Monte Porzio Catone, Rome, Italy\\
              \email{dicrisci, dantona, ventura@oa-roma.inaf.it}}

   \date{Received September ?, 2008; accepted ? ?, ?}


\abstract 
{Young populations at Z$<$Z$_{\odot}$ are being examined to 
understand the role of metallicity in the first phases of stellar evolution. 
For the analysis it is necessary to assign mass and age to Pre--Main Sequence 
(PMS) stars. While it is well known that the mass and age determination of PMS 
stars is strongly affected by the convection treatment, extending any 
calibration to metallicities different from solar one is very artificial, in 
the absence of any calibrators for the convective parameters. For solar 
abundance, Mixing Lenght Theory models have been calibrated by using the results 
of 2D radiative-hydrodynamical models (MLT-$\alpha^{2D}$), that result to be 
very similar to those computed with non-grey ATLAS9 atmosphere boundary 
condition and full spectrum of turbolence (FST) convection model both in the 
atmosphere and in the interior (NEMO--FST models).}
    {While MLT-$\alpha^{2D}$  models are not available for lower metallicities, 
    we  extend to lower Z the NEMO--FST models, in the educated guess that in such a 
    way we are simulating also at smaller Z the results of MLT-$\alpha^{2D}$ 
    models.}  
	{We use standard stellar computation techniques, in which the atmospheric
	boundary conditions are derived making use of model atmosphere grids. This
	allows to take into account the non greyness of the atmosphere, but
	adds a new parameter to the stellar structure uncertainty, namely 
	the efficiency of convection in the atmospheric structure, if convection 
	is computed in the
	atmospheric grid by a model different from the model adopted for the interior 
	integration.}
    {We present PMS models for low mass stars from 0.1 to 1.5 M$_{\odot}$ for 
    metallicities [Fe/H]= -0.5, -1.0 and -2.0. The calculations include the most 
    recent interior physics and the latest generation of non-grey atmosphere 
    models. 
    At fixed luminosity more metal poor  isochrones are hotter 
     than solar ones  by $\Delta$ $\log$ T$_{\rm eff}$/$\Delta$ $\log$Z $\sim$ 
     0.03-0.05 in the range in Z from 0.02 to 0.0002 and for ages from 10$^5$ 
     to 10$^7$ yr.}
    {}
 
\titlerunning{Updated PMS tracks at low Z for 0.1 $\le$ M/M$_{\odot}$$\le$ 1.5}
\authorrunning{Di Criscienzo et al.}
 \maketitle

%

\section{Introduction}
The study of Pre-Main Sequence (PMS) stars is important to trace the
modalities of star formation in space and time, to date young stellar systems by means of age tracers which
do not suffer the  uncertainty in the physics of the upper main
sequence stars, to derive the initial mass function of very low mass
stars and brown dwarfs, and to understand the modalities  of the stellar
rotational evolution and depletion of light elements. 
Theoretical tracks and isochrones provide an essential tool to 
understand and interpret the experimental data currently available
for young objects. Since most of these sources are located in nearby 
Galactic star forming regions, this research has been generally limited 
so far to solar chemistry or close to it \citep[e.g][]{siess2000}. 
Now that young populations at Z$<$Z$_{\odot}$ are being examined  \citep[e.g.][]{romaniello2006}, it
becomes essential to understand the role of metallicity in the first phases 
of stellar evolution, thus demanding further
investigations at lower metallicities.\\
On the other hand, the description of the PMS evolution of stars is one
of the most delicate tasks in the context of stellar astrophysics: the location
of the theoretical tracks on the Hertzprung-Russell (HR) plane depends critically
on the ingredients used to calculate the models, the equation of state and  opacities \citep[e.g.][]{dantona1993}, the boundary conditions used to match the integration of the interior with the 
atmospheric structure \citep{baraffe2002}, and the treatment of convection \citep{DM1994}.
Detailed investigations by \citet{montalban2004}   
showed that, at $T_{\rm eff}$s where atmospheric convection is present, its 
modelling plays an important role in 
determining the radius of these structures, and thus the exact excursion of 
the theoretical track on the HR diagram; this is the most critical parameter in
the computation down to T$_{\rm eff}$$\simeq$4000K. Below this temperature, both
molecular opacities and convection treatment are the dominant uncertainties.
Unfortunately, convection is a rather 
complex phenomenon, that is scarcely known from first principles, so that it 
is commonly treated by means of purely local approaches, the most popular of 
which is the Mixing Length Theory \citep[MLT]{bohm1958}, where all the physical 
uncertainties are hidden below the unique parameter $\alpha=l/H_p$, being $l$ the
mixing scale and $H_p$ the pressure scale height; among other attempts to
model convection locally, the Full Spectrum of Turbulence (FST) model by \citet[CM]{CM1991}
and \citet[CGM]{CGM1996} has been largely used in recent years.\\
A potentially powerful tool to ``calibrate" the convective model for the 
description of these evolutionary stages is the comparison between the 
theoretical loci in the HR plane and the position of PMS binaries, for which a 
rather precise estimation of the masses of the two components is possible 
\citep[e.g][]{stassun2004,boden2005}. The results obtained so far are rather 
ambiguous, and evidentiate the difficulty to find out a unique description of 
convection, holding in all cases. \citet{baraffe2002} found that several masses 
in PMS binaries demand an MLT parameter $\alpha=1.9$, but some others, lying in 
the same region of the HR diagram, require a much lower efficiency, e. g. 
$\alpha=1$. \citet{stassun2004} also found that models with 
inefficient convection in the interior should be preferred. In a case, for the binary components of V1174Ori, they found a very good agreement with the 
models by \citet{montalban2004}, employing the FST convection both in the 
interior and in the atmospheres computed by \citet{NEMO2}. 2D and 3D radiative 
hydrodynamical simulations should provide more realistic results \citep{tramp1999,ludwig2002}. Based on their 2D computations, \citet{ludwig1999} provided a 
calibration of the MLT parameter $\alpha$ for various effective temperatures 
and gravities: the meaning of this calibration is that the parameter describes 
the average efficiency of convection within the whole superadiabatic zone, in the 
region of T$_{\rm eff}$  and gravity explored by the 2D models. This 
efficiency is  not constant, but varies with the position of the star in the HR 
diagram. \citet{montalban2006} adopted such a calibration and computed the 
corresponding tracks (MLT-$\alpha^{2D}$ tracks) for solar metallicity. Exam of 
binary location with respect to the MLT-$\alpha^{2D}$ tracks  shows the same 
difficulties of previous track sets, although they result in excellent 
agreement with the components of the binary V773 Tau \citep{boden2007}. The 
MLT-$\alpha^{2D}$ tracks are not in agreement with the lithium depletion 
patterns in young clusters with the solar system abundance, but this additional 
feature can be attributed to the high metal abundance generally adopted for the solar 
model \citep{montalban2006}. The same work shows that the FST non-grey tracks 
of solar composition, with boundary conditions based on model atmospheres in 
which convection adopts the same FST model \citep{NEMO2}, present a striking 
similarity with the results of MLT-$\alpha^{2D}$ models.\\ The scope of this 
paper is to extend the computations by \citet{montalban2006} to lower 
metallicities, by naively assuming that FST models can approximate 
the results of  MLT-$\alpha^{2D}$ simulations also at lower metallicity. As we 
wish to provide a more extended set of results, while the FST model atmospheres 
are available only at T$_{\rm eff}$ $\ge$ 4000K, we adopt a way to extend the 
computation to lower T$_{\rm eff}$. We then  use MLT non grey models based on 
the atmospheric structures by \citet[AH97]{AH1997} and choose the combination of 
atmospheric and interior convection efficiency that provides results similar to the FST 
tracks at T$_{\rm eff}$ $\ge$ 4000K. 


\section{The ATON code}

The evolutionary sequences presented in the following sections were calculated 
by means of the ATON code for stellar evolution; a detailed description of the 
numerical structure of ATON can be found in \citet{ventura2007}.\\ The micro--physics adopted was recently updated, for what concerns the radiative and 
conductive opacities, and the equation of state (EOS).\\ At low temperatures, 
we use the latest release of the opacity tables by \citet{ferguson2005}, 
completed for $T\geq 10000$K by the OPAL opacities, in the version documented 
by \citet{iglesias1996}. \\
The EOS adopted in most of the $T-P$ plane is the latest release of the OPAL 
EOS\footnote{\texttt{http://physci.llnl.gov/Research/OPAL/EOS$\_$2005/}}, overwritten in the pressure ionization regime by the 
EOS from \citet{saumon1995}, and extended to the high-density, high temperature 
domain according to the treatment by \citet{stoltzmann2000}.\\ The formal 
borders of the convective regions were found by means of the classic 
Schwartzschild criterium. The FST scheme (CM or CGM) was used to determine the 
temperature gradient in zones unstable to convection. Nuclear burning within 
convective regions was treated according to the instantaneous mixing 
approximation.\\
We computed models for three metallicities [Fe/H]= -0.5,-1.0 and 2.0 with an adopted helium mass fraction Y=0.25. The solar metallicity is assumed to be Z=0.02 and the solar mixture for opacities and EOS is taken from \citet{grevesse}. In this work we adopted solar-scaled mixtures in all cases, though $\alpha-$enhanced mixtures are also available.\\

 \subsection{Atmospheric structure and boundary conditions} 
In the atmosphere, the opacities have a strong dependence on the frequency, 
so that an atmospheric integration based on a Rosseland--type average 
or on an approximate T($\tau$) relation, like in a grey atmosphere, is not
adequate for the description of the structure. Often this problem is attacked
by adopting the stratification temperature vs. pressure by integrating appropriate
model atmospheres, as for example by means of the Kurucz code \citep{kurucz1998}.
Below $\sim$4000K, the role of triatomic molecules becomes important, and the most
adequate model atmospheres are so far the models by \cite{AH1997}.
Generally, the boundary conditions for the interior structure computation are then
the physical quantities deriving from such integration down to a fixed value of the
optical depth $\tau$, namely $\tau_{\rm ph}$. Tables of boundary conditions at the chosen
$\tau_{\rm ph}$ are used to derive the stellar gravity and T$_{\rm eff}$ by interpolation.
This procedure hides a problem: the stellar convection in the atmosphere is computed
by assuming an efficiency of convection. In the MLT model, for example, the grids of
model atmospheres are computed by fixing the $\alpha$ parameter to a value $\alpha_{\rm atm}$.
For the interior computation, it may be necessary to adopt a different $\alpha$. For example,
in order to fit the solar model (that is, to obtain the solar radius at the solar age),
a value $\alpha_{\rm int}$=1.9 is used by \citet{baraffe2002}, while the AH97 grid adopted for the
atmospheric integration has $\alpha_{\rm atm}$=1. As the MLT must be generally understood as a
way of obtaining an ``average" efficiency of convection, more than a model that
allows to derive the correct temperature atmospheric structure, this problem should not
worry too much, but we should remember that in this way we are introducing 
a dependence of the structure on another parameter, namely $\tau_{\rm ph}$. This problem is fully
discussed by \citet{montalban2004}.\\
\citet{NEMO2} made available
grids computed by means of an improved version of Kurucz code \citep[NEMO;][]{kurucz1998, castelli1997}. 
They considered both MLT models with $\alpha_{\rm atm}$=0.5, FST models following CM 
and FST models according to CGM. If FST convection is adopted both in the atmospheric grid 
and in the interior, the model computation does not show temperature gradients discontinuities
\citep{montalban2004}.\\
\begin{figure*}
\centering
\includegraphics[angle=0,scale=.7]{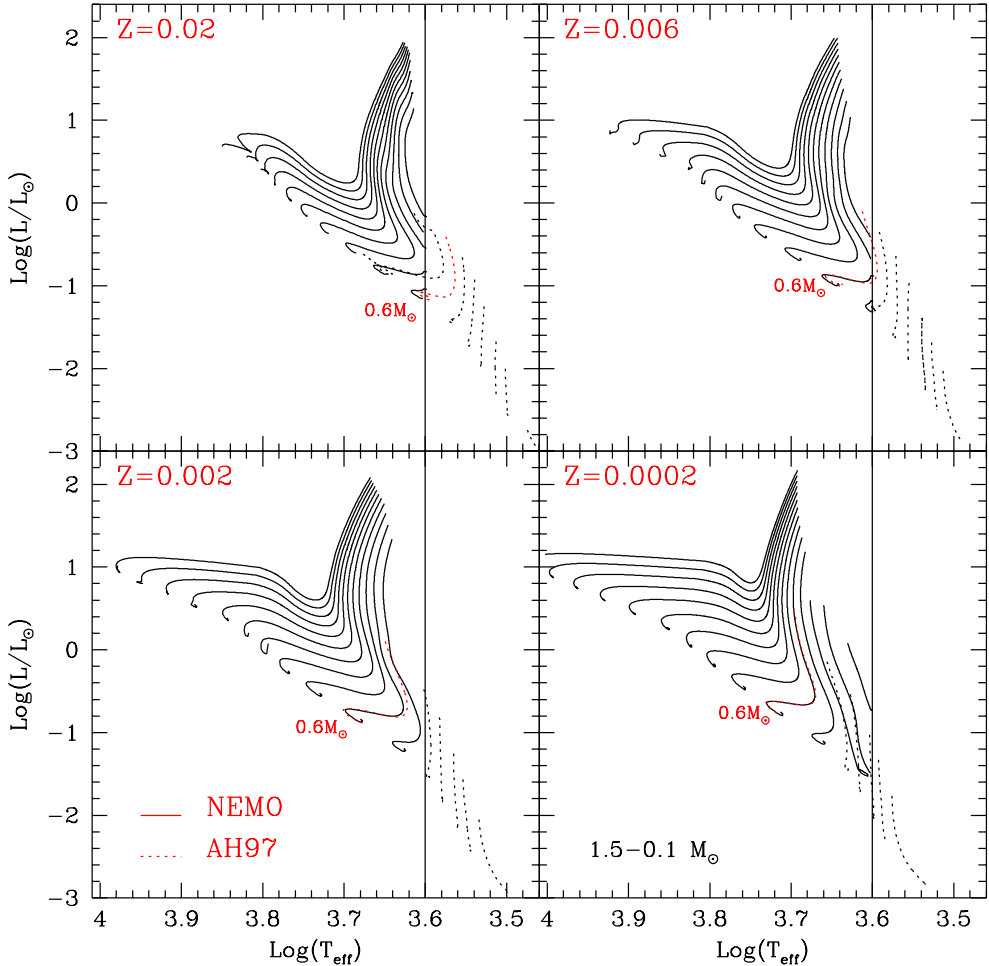} 
\caption{Evolutionary tracks for the labelled metallicities and for masses 
0.1, 0.15, 0.2, 0.3, 0.4, 0.5, 0.6, 0.7, 0.8, 0.9, 1.0, 1.1, 1.2, 1.3, 1.4, 1.5 M$_{\odot}$.  
The solid ones are calculated using NEMO grids of atmosphere models  
and FST treatment of convection in the interior, while dashed lines are MLT-tracks 
$\alpha_{\rm int}$=2.0, $\tau_{\rm ph}$=3) obtained using AH97 models of atmospheres. To facilitate the reading of this figure  in 
each panel the evolutionary track corresponding to 0.6 M$_{\odot}$ 
is labelled; for this value of mass both the MLT--AH97 and FST--NEMO tracks are reported (see also Fig. \ref{figALFA}).} 
\label{fig1}
\end{figure*} 
Different sets of boundary conditions were considered in this work: FST--NEMO 
grids (with CGM convection) and the AH97 grids. 
The boundary conditions (BCs) at fixed values of photospheric 
optical depth   $\tau_{\rm ph}$ contain, for each T$_{\rm eff}$, gravity and metallicity, 
the temperature, pressure and geometrical depth at the 
chosen $\tau_{\rm ph}$. The latter quantity is the difference
between the radius where convection begins and 
$\nabla_{rad}$ becomes larger than $\nabla_{ad}$ and the radius
at $\tau_{\rm ph}$, a quantity necessary for the
computation of the FST fluxes. 
As suggested by Heiter et al. (2002), we choose  
$\tau_{\rm ph}$=10,  to avoid discrepancies in the physical quantities 
due to the turbulent pressure 
(not included in the atmosphere modeling but present in the integration of the interior) and to reduce the differences due to different 
opacity tables used at both sides of $\tau_{\rm ph}$. The boundary 
condition for the internal structure  are determined  by spline interpolation of these tables. From initial T$_{\rm eff}$  and $\log$ L/L$_{\odot}$ we 
determine P and T  at the last point of the internal structure ($\tau_{\rm ph}$). 
An iterative procedure is performed until the P($\tau_{\rm ph}$) and 
T($\tau_{\rm ph}$) values derived from the interior and from the atmosphere 
models converge. The T$_{\rm eff}$ range of these grids is between 
4000-10000 K, at lower T$_{\rm eff}$  we have used 
AH97 models, which include the 
contribution of many more molecular lines dominating the opacities with the respect to ATLAS9. In the AH97 models, convection is 
treated with MLT and $\alpha_{\rm atm}$=1. For metallicities lower than sola,r the 
available models have 3000$\le$T$_{\rm eff}$$\le$10000 K and surface gravity 
from $\log${\it g}=3.5 to 6.0.
\begin{figure}
\centering
\includegraphics[angle=0,scale=.45]{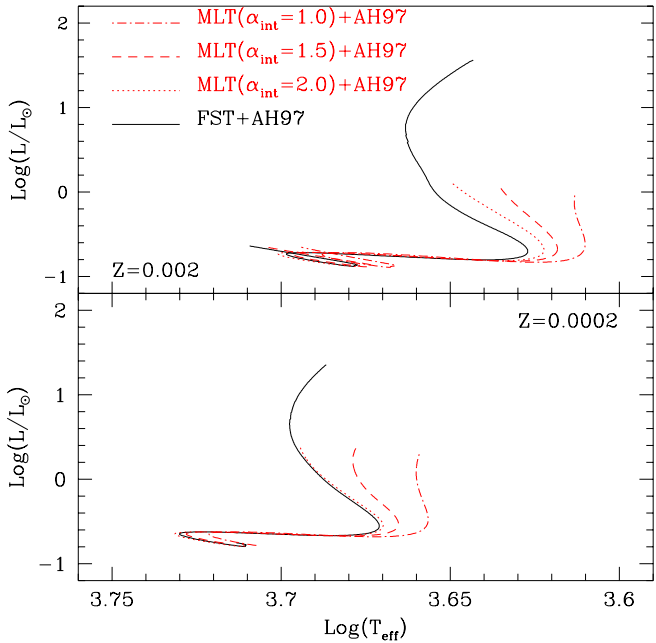} 
\caption{Non-grey evolutionary tracks for 0.6 M$_{\odot}$ and Z=0.002 
({\it upper panel}) and Z=0.0002 ({\it bottom panel}) . The solid line is FST 
tracks with NEMO-FST atmospheres with $\tau$$_{\rm ph}$=10. The others are   
MLT tracks obtained using the atmosphere models by AH97 with different values 
of $\alpha_{\rm int}$\ in the computation of the sub-atmospheric convection and with the 
match point at $\tau_{\rm ph}$=3.} 
\label{figALFA}
\end{figure}

\section{The evolutionary tracks}

In Fig. \ref{fig1} we present evolutionary tracks computed with non grey 
atmospheres for masses in the range 0.1 $\le$ M/M$_{\odot}$ $\le$ 1.5, and for 
metallicities [Fe/H]= 0.0, -0.5, -1.0, -2.0. \\
In the top-left panel the solar tracks at M $\ge$ 0.6 M$_{\odot}$ are directly taken by previous  computation by \citet{montalban2004}, which were performed using  the NEMO--FST grids by \citet{NEMO2} available only for T$_{\rm eff}$$\ge$4000 K.
The comparison of these solar metallicities tracks with other models present in the literature, in particular with those of Siess et al. 2000, was made by Montalban et al. 2004 .\\ However \citet{montalban2006} have shown that these models are very similar to the MLT-$\alpha^{2D}$ models; here we extend the computation to 
lower metallicities. \\
Since measurements suggest a primordial abundance of 
deuterium  $\sim$ 3.0--5.0 $\cdot$ 10$^{-5}$ \citep[e.g.][]{tosi1996},  
we fix the initial deuterium  abundance at X$_D$=4  $\cdot$ 10$^{-5}$ 
by mass, fairly representative of the D-abundance of the Population II stars. A 
value X$_D$=2 $\cdot$ 10$^{-5}$ is often adopted for  models with solar 
abundances \citep{siess2000}. 
\\ NEMO models are {\bf available} only for T$_{\rm eff}$$\ge$4000K. 
Within the context of non grey modelling, the only atmospheres 
available at lower $T_{\rm eff}$ are those by \citet{AH1997}, that 
use the MLT treatment of convection. We therefore decided to extend our models 
to lower masses  by using these atmosphere's models. 
  The same convection model is adopted in the  interior computation, and the
 match  with atmospheres is done at $\tau_{\rm ph}$=10
 in the first case and $\tau_{\rm ph}$=3 in the second one.
The free parameter $\alpha_{\rm int}$ was set in order to provide a reasonable continuity 
between the FST tracks and the MLT ones at temperatures just exceeding 4000K. 
Fig \ref{figALFA} 
shows that  $\alpha_{\rm int}=2$ is a reasonable choice especially at lower 
metallicity (see also the evolutionary tracks for M=0.6 M$_{\odot}$ in 
Fig.\ref{fig1}). For these MLT  models we choose $\tau_{\rm ph}$=3 which is a 
good compromise since  we avoid having a large influence from $\alpha_{\rm atm}$
\citep{montalban2004}.    The evolutionary tracks computed with the MLT 
treatment of convection are indicated with dashed lines in Fig.\ref{fig1}; 
since the AH97 atmospheres are available only in the range of gravities $\log$ 
{\it g} $\ge$ 3.5, we had to skip the deuterium burning phase from our 
computations. \\
\begin{figure}
\centering
\includegraphics[angle=0,scale=.45]{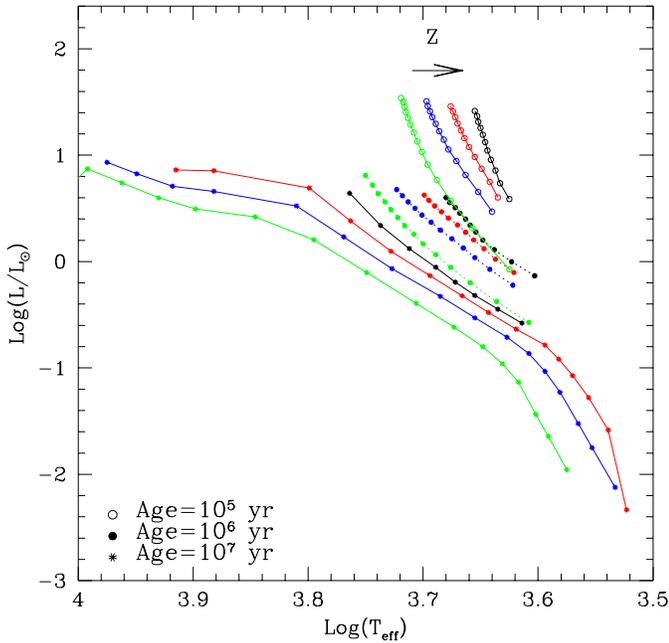} 
\caption{Comparison between the isochrones calculated for 10$^5$, 10$^6$, 10$^7$ yr for different metallicities (from left to right: [Fe/H]=-2.0, -1.0, -0.5, 0)  } 
\label{fig2}
\end{figure}
As outlined in the introduction, PMS stars can be very useful age tracers. 
Comparing the theoretical results with the observed loci of stellar 
sources in associations requires the computation of the isochrones.
In Fig.\ref{fig2} we report three different groups of isochrones 
(corresponding to 10$^5$, 10$^6$, 10$^7$yr) for each of the metallicities investigated. \\ The youngest isochrones do not include the low mass stars, because, as already stressed, we started the evolution after the deuterium burning phase.
As for very low mass models (those with AH97 atmospheres) we skip the
D burning phase, the ages of these PMS stars must be taken with
caution. At $10^7$yr, for example, including  D burning would have
increased the age by about only  3$\%$ at 0.5 M$_{\odot}$, but by about
30$\%$ at  0.1M$_{\odot}$  \citep{DM1997}.
Young ages as low as $10^5$yr are also uncertain, because of the unsure
role played by the protostellar accretion phase \citep{palla}.
{\bf It} is evident from the figure that at fixed age and luminosity more metal poor isochrones are hotter by an amount which is almost independent of luminosity  and that these amounts became lower at lower metallicity. 
The effect is mostly due to 
the opacity reduction, as it is well known in the study of main sequence 
models.

 \section{Conclusions} 
A grid of stellar evolutionary tracks for low metallicity Pre--Main Sequence 
stars with masses between 0.1 and 1.5 M$_{\odot}$ was presented. 
These models are 
based on up-to-date physics and updated non grey atmosphere models were used. A 
coherent treatment of convection in the interior and exterior region of the 
star was employed at  T$_{\rm eff}$$\ge$4000K. We extended our computations to 
models of smaller masses by using the AH97 grid of model atmospheres. The 
parameters $\tau_{\rm ph}$ and $\alpha_{\rm int}$ were  chosen to provide a smooth 
transition between the two model sets. This, of course, is only an educated 
guess to the problem of Pre--Main Sequence models at low metallicity. 
\\ The models (available in electronic form at the WEB location 
\texttt{http://www.mporzio.astro.it/$\%$7Etsa/}) can now be confronted to the 
complex realm of very young objects, providing important information on ages and 
star formation processes, and, on the other hand, providing some 
new constraints for PMS models.

\begin{acknowledgements}
It is a pleasure to thank J. Montalban for having provided 
the tables of boundary conditions  for low metallicities and for precious comments on the manuscript. 
Financial support for this study was provided by MIUR under the PRIN project 
``Asteroseismology: a necessary tool for the advancement in the study 
of stellar structure, dynamics and evolution'', P.I. L. Patern\'o.\\ 
\end{acknowledgements}

\end{document}